# Uniform Spin Qubit Devices in an *All-Silicon* 300 mm Integrated Process

N. I. Dumoulin Stuyck[1,2], R. Li[1], C. Godfrin[1], A. Elsayed[1,3], S. Kubicek[1], J. Jussot[1], B. T. Chan[1],
F. A. Mohiyaddin[1], M. Shehata[1,3], G. Simion[1], Y. Canvel[1], L. Goux[1],
M. Heyns[1,2], B. Govoreanu[1] & I. P. Radu[1]

[1]Imec, B-3001 Leuven, Belgium, email: nard.dumoulin@imec.be
[2]Department of Materials Engineering (MTM), KU Leuven, B-3001 Leuven, Belgium
[3]Department of Physics and Astronomy, KU Leuven, B-3001 Leuven, Belgium

**Abstract**

Larger arrays of electron spin qubits require radical improvements in fabrication and device uniformity. Here we demonstrate excellent qubit device uniformity and tunability from 300K down to mK temperatures. This is achieved, for the first time, by integrating an overlapping polycrystalline silicon-based gate stack in an 'all-Silicon' and lithographically flexible 300mm flow. Low-disorder Si/SiO$_2$ is proved by a 10K Hall mobility of $1.5 \cdot 10^4$ cm$^2$/Vs. Well-controlled sensors with low charge noise (3.6 µeV/√Hz at 1 Hz) are used for charge sensing down to the last electron. We demonstrate excellent and reproducible interdot coupling control over nearly 2 decades (2-100 GHz). We show spin manipulation and single-shot spin readout, extracting a valley splitting energy of around 150 µeV. These low-disorder, uniform qubit devices and 300mm fab integration pave the way for fast scale-up to large quantum processors.

**Keywords:** Si-MOS quantum dots, polycristalline silicon, spin qubits, 300 mm quantum computing.

## Introduction

Spin states of confined electrons have promising properties as quantum bits or qubits. To this goal, different approaches include Si-MOS [1,2], Si/SiGe [3] and SOI-based technologies [4], in either bulk or nanowire technologies. However, scaling up to large spin qubit arrays is still a grand challenge. One of the most important limitations is disorder from fabrication, i.e. interface defects. Of particular importance is a uniform and symmetric tunnel barrier control [5]. Here, we demonstrate multiple qubit device initialization, manipulation, and read-out, with uniform qubit behavior and low disorder structures.

## Device Fabrication

Qubit devices are fabricated using a 300 mm integration flow which allows flexible design implementations [1]. We have reported prior on a similar flow but with a TiN gate-stack [2]. Here, TiN gates are replaced by a novel overlapping polycrystalline silicon gate-stack. This **'all-Silicon'** approach aims to increase thermal compatibility at cryogenic temperatures, reducing strain associated disorder [6]. The triple gate layer stack is formed on a thermally grown SiO$_2$ film (8 nm thick) with 30nm-tick gates, isolated by ALD-based SiO$_2$ (5 nm). Wafer-scale process monitoring shows **tight CD control for all gate layers** (**Fig.1**). Process sanity is further tested electrically using specific monitors. As a basic process uniformity metric, all gate levels show uniform V$_{th}$ distributions collected on test transistors (**Fig.2a,b**).

## Results

To assess interface quality, Hall bar mobility measurements are performed at a temperature of 10K. *Very large mobilities of up to 15000 cm$^2$/Vs confirm high interface quality* (**Fig. 2c**). Wafer-scale Single Electron Transistor (SET) IdVg curves at 300K show reproducible channel control for all gates, with *excellent multi-gate V$_{th}$ uniformity down to 8 mK* as shown by 3 randomly selected devices (**Fig.3**). The narrow V$_{th}$ distribution and smooth subthreshold turn-on indicate a high quality of the nanostructure. This is further confirmed by successful formation of an electron sensing island between the left (LB) and right (RB) barrier gates, (**Fig.4,5a**). Noise characterization returns an expected *1/f noise power density spectrum with a low amplitude of 3.6 μeV/√Hz at 1 Hz* (**Fig.5b,c**). SET sensing measurements verify device operation in single electron occupation for both the single- and double quantum dot mode (**Fig.6,7**). In the low electron-count regime, dynamic control of the tunnel coupling t$_c$ is crucial for high fidelity two-qubit gates. Coupling strength t$_c$ is extracted from the charge transition width, using a *lever arm of 0.3 eV/V* derived from the thermal broadening (**Fig.8a,b**) [7]. We show *reproducible and symmetric t$_c$ control of up to 100 GHz, over nearly 2 decades*, using the barrier gate voltage V$_{B1}$ for two different charge transitions: (0,1)-(1,0) and (1,1)-(2,0) (**Fig.8c,d**). The excellent tunability and high t$_c$ achieved in the single electron regime further justify the exceptional interface quality, as required for high-fidelity qubits. For qubit operation, a static magnetic field B$_0$ is applied, splitting the electron spin states by the Zeeman energy E$_Z$. The spin of the last electron in the dot can be read out using a spin-to-charge conversion technique [8]. Spin lifetime T$_1$ is extracted and shows an expected B$_0$ dependency [9]. A relaxation hotspot, occurring when E$_Z$ matches the valley splitting E$_V$, is measured for fields between 1.1 T and 1.3 T, corresponding to *E$_V$ in the range of 127 to 150 μeV*. Spin states are manipulated using Electron Spin Resonance (ESR) techniques with an on-chip antenna (**Fig.9b**). The extracted FWHM of ~2 MHz is consistent with that of donor bound electron spins in natural Si [10].

## Conclusion & Outlook

We use a novel 'all-Silicon' integration approach to demonstrate excellent Si-MOS spin qubit devices on natural-Si substates. Well-controlled process uniformity and high Si/SiO$_2$ interface quality lead to reproducible electrical response across the entire temperature range (300K-mK). We demonstrate excellent tunnel coupling control in last electron regimes. This unprecedented reproducibility opens the path towards large-scale, spin-based quantum processors.

## Acknowledgements

This work was performed as part of imec's Industrial Affiliation Program on Quantum Computing. N. I. Dumoulin Stuyck gratefully acknowledges FWO-Vlaanderen for a Strategic Basic Research PhD fellowship, application number 1S60020N.

## References

[1] B. Govoreanu *et al*, *SNW* (2019). [2] R. Li *et al*, *IEDM* (2019). [3] R. Pillarisetty *et al*, *IEDM.* (2019). [4] R. Maurand *et al*, *Nat. Comm.* (2016). [5] M.D. Reed *et al*, *PRL.* (2016). [6] F. A. Mohiyaddin *et al*, *IEDM.* (2019). [7] T. Hensgens *et al, Nat. (2017).* [8] J.M. Elzerman *et al, Nat.* (2004). [9] C. H. Yang *et al, Nat. Com.* (2013). [10] J. J. Pla *et al*, *Nat.* (2012).

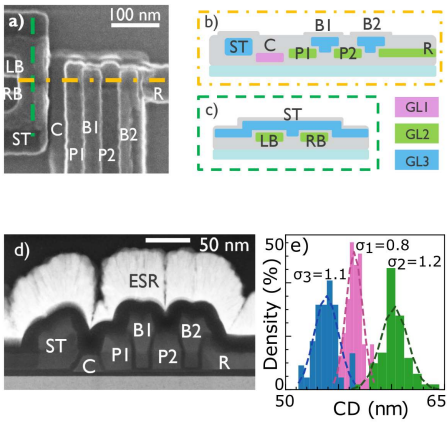
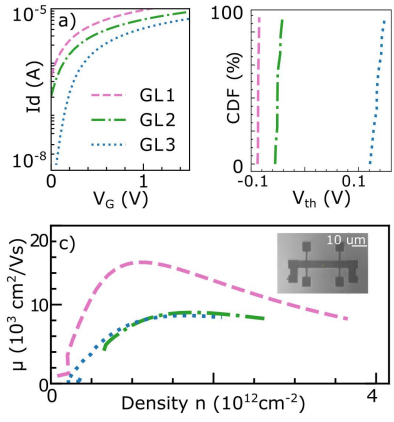
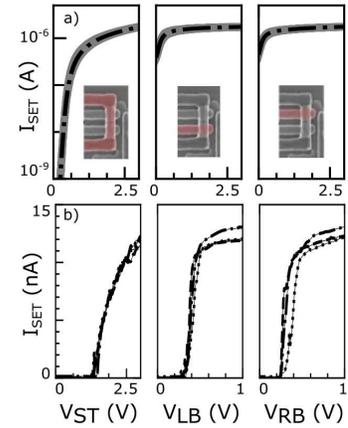

**Fig. 1.** a) Device SEM. b) and c) Schematics of qubit structure and SET, respectively. d) Device TEM along b). e) CD-SEM gate width statistics (in nm) for gate levels 1, 2 & 3. The actual device physical width (TEM) is smaller than the in-line CD monitors.

**Fig. 2.** a) 10 μm channel transistor IdVg. b) corresponding $V_{th}$ distribution. c) Hall bar mobility taken at 10K for gate layer 1, 2 and 3. Inset: SEM of the Hall bar structure.

**Fig. 3.** a) Wafer scale SET IdVg at 300K for different gates indicated in inset figures. Non sweeping gates are kept at high voltages. b) Same as a) but taken at 8 mK for 3 randomly selected devices.

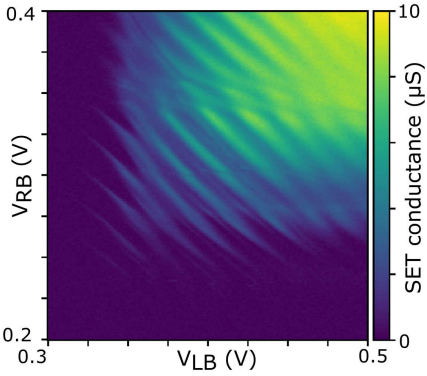
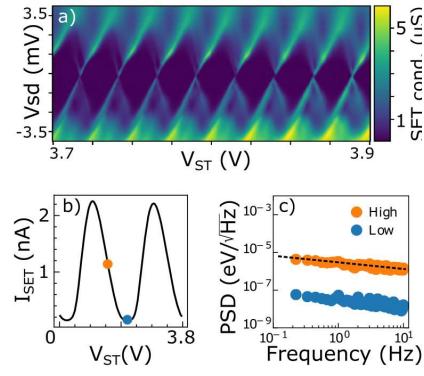
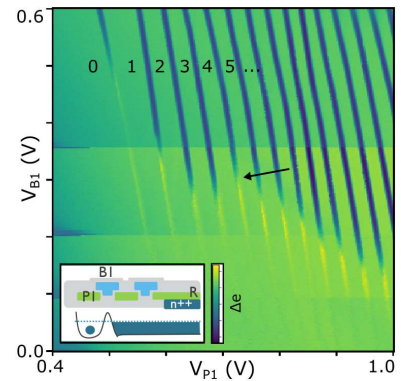

**Fig. 4.** The SET conductance measured as a function of $V_{LB}$ and $V_{RB}$ shows clean diagonal lines with a 45° slope, corresponding to a well-defined quantum dot between the two barrier gates.

**Fig. 5.** a) SET Coulomb diamonds over large $V_{ST}$ span. b) Line cut across a) at $V_{sd}$ = 1mV. c) SET noise power spectral density taken at a high and low sensitivity point, showing 1/f dependence.

**Fig. 6.** Single quantum dot stability map. No further transitions can be observed at lower voltages, confirming the last electron regime. Arrow highlights the orbital shell filling feature on the 4$^{th}$ to 5$^{th}$ electron occupation transition. Inset: conduction band energy & color bar.

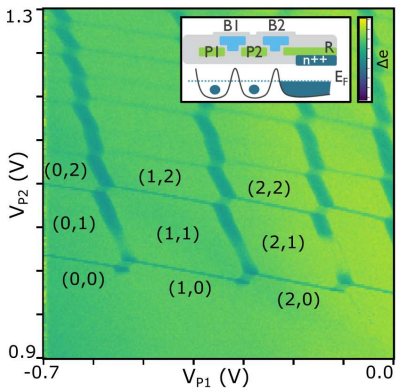
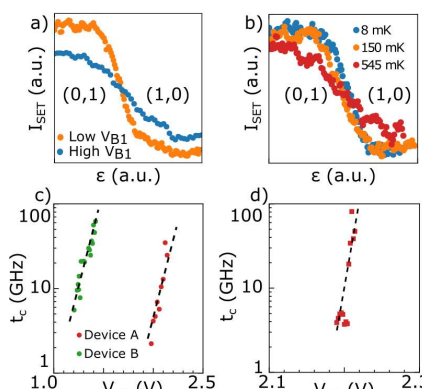
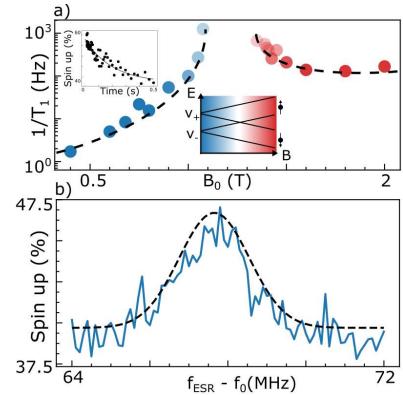

**Fig. 7.** Double quantum dot stability map. ($N_1$, $N_2$) denote the electron occupation in dot 1 and 2, respectively. No further transitions are observed at lower gate voltages, confirming the last electron occupation in both dots. Inset: conduction band energy & color bar.

**Fig. 8.** a) SET response for the DQD (0,1)-(1,0) transition for different barrier height across the detuning axis ε, $t_c$ is extracted from the transition width. b) Temperature broadening, used to extract the lever arm. c) $t_c$ control at the (1,0)-(0,1) transition. d) $t_c$ control at (1,1)-(0,2).

**Fig. 9.** a) Spin lifetime as function of $B_0$. When the Zeeman energy equals the valley splitting a relaxation hotspot can be resolved. Inset: energy spectrum & spin decay trace for 0.6 T. b) Spin resonance with the on-chip ESR antenna. The resonance frequency is $f_0$ ~ 19 GHz with FWHM ~ 2 MHz.